\documentclass[twocolumn,showpacs,preprintnumbers,prl]{revtex4}
\usepackage{graphicx}
\usepackage{graphicx}
\usepackage{epsfig}

\begin{document}

\title{Quantum criticality and universal scaling of strongly attractive
spin-imbalanced Fermi gases in a 1D harmonic trap}
\author{Xiangguo Yin$^{1}$, Xi-Wen Guan$^{2,\ddagger}$, Shu Chen$^{1}$ and
Murray T Batchelor$^{2,3}$}
\affiliation{${1}$ Institute of Physics, Chinese Academy of Sciences, Beijing 100190,
China}
\affiliation{${2} $ Department of Theoretical Physics, Research School of Physics and
Engineering, Australian National University, Canberra ACT 0200, Australia}
\affiliation{${3}$ Mathematical Sciences Institute, Australian National University,
Canberra ACT 0200, Australia}

\begin{abstract}
We investigate quantum criticality and universal scaling of strongly
attractive Fermi gases confined in a one-dimensional (1D) harmonic trap. 
We demonstrate from the power-law scaling of the thermodynamic properties that 
current experiments on this system 
are capable of measuring universal features at quantum criticality, 
such as universal scaling and Tomonaga-Luttinger liquid physics. 
The results also provide insights on recent measurements of key features of the 
phase diagram of a spin-imbalanced
atomic Fermi gas [Y. Liao \emph{et al.}, Nature \textbf{467}, 567 (2010)] and
point to further study of quantum critical phenomena in ultracold atomic
Fermi gases.
\end{abstract}

\date{\today }
\pacs{03.75.Ss, 03.75.Hh, 02.30.Ik, 05.30.Rt}
\maketitle

Very recently, the one-dimensional (1D) strongly attractive two-component
Fermi gas has attracted much attention from theorists 
\cite{Orso,HuiHu,Guan,Mueller,Kakashvili,Wadati,Angela} and experimentalists 
\cite{Liao} due to the existence of an exotic pairing mechanism. Investigation
\cite{YangK,Feiguin,Liu} shows that this novel pairing is closely related to
the elusive Fulde-Ferrel-Larkin-Ovchinnikov (FFLO) \cite{FFLO} state
involving BCS pairs with nonzero center-of-mass momenta.

The 1D Fermi gas is one of the most important exactly solvable quantum
many-body systems. It was solved long ago by Yang \cite{Yang} and Gaudin
\cite{Gaudin} using the Bethe ansatz. Although the study of the attractive
Fermi gas was initiated soon after \cite{Yang-a, Takahashi-a}, it was not
until much later that this model began to receive more attention \cite{Fuchs}. 
In terms of the polarization $p$ the model exhibits three quantum phases
at zero temperature \cite{Orso,HuiHu,Guan}: the fully paired phase which is
a quasi-condensate with $P=0$, the fully polarized normal Fermi liquid with 
$P = 1$, and the partially polarized (1D FFLO-like) phase for $0 < P < 1$. 

\begin{figure}[t]
{{\includegraphics [width=1.0\linewidth]{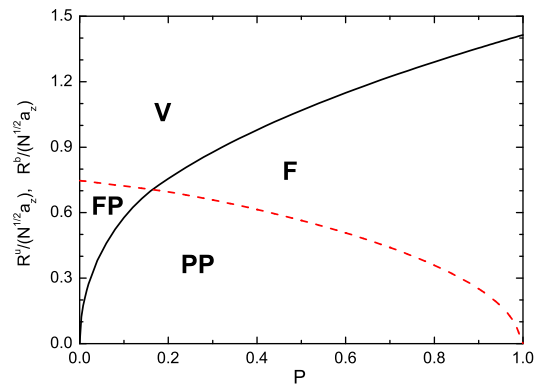}}}
\caption{Phase diagram for two-component fermions in an
harmonic trap at zero temperature as a function of polarization with $%
Na_{1D}^2/a_z^2=0.2$. Here $FP$ denotes the fully paired phase, $F$ the
fully unpaired phase and $PP$ the FFLO-like phase. $V$ is the vacuum phase.}
\label{fig:phase}
\end{figure}

For a trapped imbalanced Fermi gas it is found \cite{HuiHu,Orso} within the
local density approximation (LDA) that a partially polarized 1D FFLO-like
state sits in the trapping center surrounded by wings composed of either a
fully paired state or a fully polarized Fermi gas. The phase boundaries of
the zero-temperature phase diagram shown in Fig.~\ref{fig:phase} can be
determined precisely from the exact solution by the vanishing of the axial
density difference (solid line) and the minority state axial density (dashed
line). The key features of this $T=0$ phase diagram were experimentally
confirmed using finite temperature density profiles of trapped fermionic 
${}^6$Li atoms \cite{Liao}.

Most recently, quantum criticality and universal scaling
behaviour \cite{QC-book} are being explored in 
low-dimensional cold atomic matter. For example, in experiments on 
the 2D Bose gas \cite{2D-Bose}, following a 
theoretical scheme for mapping out quantum criticality \cite{Zhou-Ho,Mueller2}. 
From this viewpoint, the 1D imbalanced Fermi gas, 
exhibiting novel phase transitions at $T=0$, is particularly valuable to
test universal scaling through finite temperature density profiles of
trapped ultracold atoms \cite{Guan-Ho}.
Here we illustrate that finite-temperature properties of the quasi-1D
trapped Fermi gas allow the exploration of a wide range of physical
phenomena, such as universal Tomonaga-Luttinger liquid (TLL) physics,
scaling theory and the nature of the FFLO state at quantum criticality. %

\textit{The Model and equation of state.-} We consider the 1D $\delta$%
-interacting attractive spin-1/2 Fermi gas with $N=N_{\uparrow
}+N_{\downarrow}$ fermions of mass $m$ and external magnetic field $H$. The
system is described by the hamiltonian
\begin{eqnarray}
\mathcal{H} &=&-\frac{\hbar ^{2}}{2m}\sum_{i=1}^{N}\frac{\partial ^{2}}{
\partial x_{i}^{2}}+g_{1D}\sum_{i=1}^{N_{\uparrow
}}\sum_{j=1}^{N_{\downarrow }}\delta \left( x_{i}-x_{j}\right)  \nonumber \\
&&-\frac{1}{2}H\left( N_{\uparrow }-N_{\downarrow }\right)
\label{Hamiltonian}
\end{eqnarray}
in which the three terms are kinetic energy, interaction energy and Zeeman
energy, respectively. Here the inter-component interaction is related to an
effective 1D scattering length $g_{1D}=-\frac{2\hbar^2}{ma_{1D}}$ which can
be tuned from the weakly interacting regime ($g_{1D}\rightarrow 0^-$) to the
strong coupling regime ($g_{1D}\rightarrow -\infty $) via Feshbach
resonances and optical confinement. For convenience, we define the
interaction strength as $c=mg_{1D}/\hbar ^{2}$ and dimensionless parameter $%
\gamma =c/n$ for physical analysis, where $n=N/L$ is the linear density and $%
L$ is the length for system. We set Boltzmann constant $k_{B}=1$.

For the strongly attractive spin-1/2 Fermi gas at finite temperatures, the
thermodynamics of the homogeneous system is described by two coupled Fermi
gases of bound pairs and excess fermions in the charge sector and
ferromagnetic spin-spin interaction in the spin sector. Spin fluctuations
are suppressed by a strong effective magnetic field at low temperatures. For
the physically interesting low temperature and strong coupling regime, i.e.,
$T\ll \epsilon _{b},H\text{ \ and \ }\gamma \gg 1$ a high precision equation
of state \cite{Zhao,Guan-Ho,GB} can be derived from the thermodynamic Bethe
ansatz (TBA) equations \cite{Takahashi} in terms of the Yang-Yang grand
canonical ensemble \cite{Y-Y}. Using the binding energy $\varepsilon
_{b}=\hbar ^{2}c^{2}/4m$ as the unit of energy and defining $\tilde{\mu}=\mu
/\varepsilon _{b}$, $h=H/\varepsilon _{b}$, $t=T/\varepsilon _{b}$, $\tilde{p%
}=p/\left\vert c\varepsilon _{b}\right\vert $, the pressure $\tilde{p}=%
\tilde{p}^{b}+\tilde{p}^{u}$ of the system is found to be
\begin{eqnarray}
\tilde{p}^{b} &=&-\frac{t^{3/2}f_{3/2}^{b}}{2\sqrt{\pi }}\left( 1-\frac{%
t^{3/2}f_{3/2}^{b}}{16\sqrt{\pi }}-\frac{t^{3/2}f_{3/2}^{u}}{\sqrt{2\pi }}%
\right)   \nonumber \\
\tilde{p}^{u} &=&-\frac{t^{3/2}f_{3/2}^{u}}{2\sqrt{2\pi }}\left( 1-\frac{%
t^{3/2}f_{3/2}^{b}}{\sqrt{\pi }}\right)   \label{Ep}
\end{eqnarray}%
where $f_{n}^{b}=\mathrm{Li}_{n}\left(-e^{A_{b}/t}\right) $ and $f_{n}^{u}=%
\mathrm{Li}_{n}\left(-e^{A_{u}/t}\right) $ in terms of the standard polylog
function $\mathrm{Li}_{n}\left( x\right) $, with %
\begin{eqnarray}
A_{b}&=&2\tilde{\mu}+1-\tilde{p}^{b}-4\tilde{p}^{u}-\frac{t^{5/2}f_{5/2}^{b}}{
16\sqrt{\pi }}-\sqrt{\frac{2}{\pi }}t^{5/2}f_{5/2}^{u}\nonumber  \\
A_{u}&=&\tilde{\mu}+\frac{h}{2}-2\tilde{p}^{b}-\frac{t^{5/2}f_{5/2}^{b}}{2\sqrt{\pi
}}+f_{s} \label{chemical2}
\end{eqnarray}
Here $f_{s}=te^{-h/t}e^{-2\tilde{p}^{u}/t}I_{0}\left( 2\tilde{p}%
^{u}/t\right) $ and $I_{n}\left( x\right) =\sum_{k=0}^{\infty }\frac{1}{%
k!\left( n+k\right) !}\left( x/2\right) ^{n+2k}$. The thermodynamic
quantities such as the particle density $n$, $n^{u}$ for unpaired fermions, $%
n^{b}$ for paired fermions, and compressibility
follow from Eq. (\ref{Ep}) and the standard thermodynamic
relations. The total pressure serves as the equation of state
which provides high precision thermodynamics
over a wide temperature range $T<0.2\varepsilon _{b}$, see \cite%
{Guan-Ho,Zhao}.

\textit{Quantum phase diagram in a harmonic trap.-} For spin imbalanced
attractive fermions in an harmonic trap, the equation of state (\ref{Ep})
can be reformulated within the LDA by the replacement $\mu \left( z\right)
=\mu \left( 0\right) -\frac{1}{2}m\omega _{z}^{2}z^{2}$ in which $z$ is the
position and $\omega _{z}$ is the frequency within the trap. Using the
dimensionless chemical potential this becomes $\tilde{\mu}\left( z\right) =%
\tilde{\mu}\left( 0\right) -2\tilde{z}^{2}$, where $\tilde{z}=z/\left(
a_{z}^{2}\left\vert c\right\vert \right) $ with the harmonic characteristic
radius $a_{z}=\sqrt{\hbar /\left( m\omega _{z}\right) }$. The total particle
number $N=\int_{-\infty }^{\infty }n\left( z\right) dz$ and polarization $%
P=\int_{-\infty }^{\infty }n^{u}\left( z\right) dz/N$ become
\begin{eqnarray}
{Na_{1D}^{2}}/{a_{z}^{2}} &=&4\int_{-\infty }^{\infty }\tilde{n}\left(
z\right) d\tilde{z}  \nonumber \\
P &=&4\int_{-\infty }^{\infty }\tilde{n}^{u}\left( z\right) d\tilde{z}\times
{a_{z}^{2}/}\left( {Na_{1D}^{2}}\right)  \label{N_P}
\end{eqnarray}%
where $\tilde{n}\left( z\right) =1/\left\vert \gamma \left( z\right)
\right\vert $ and $\tilde{n}^{u}\left( z\right) =n^{u}\left( z\right) /|c|$.

At finite temperatures, the phase boundaries can be determined from the
equation of state (\ref{Ep}) within the LDA (\ref{N_P}), where the
boundaries of vanishing density difference (black solid line) and vanishing
unpaired fermions (red dashed line) form three phases (see Fig.~\ref%
{fig:phase}). As $t \rightarrow 0$, the phase boundaries determined via (\ref%
{Ep}) are consistent with the zero-temperature results. For strong attraction 
with polarization, the atoms with opposite spin
states form hard-core bosons which are strongly correlated with excess
fermions. The polarization can be changed by tuning the effective magnetic
field. This results in two extreme phases -- the fully paired and fully
polarized phases. At zero temperature, these phases intersect at global
polarization $P=P_c$, where we find
\begin{equation}
P_c=\frac{1}{5}\left(1-\frac{64}{75|\gamma|}\right)+O\left(1/\gamma^2\right).
\end{equation}
Here the phase boundaries are sharp lines and $P_c$ is the intersection
point in Fig.~\ref{fig:phase}. At finite temperatures the intersection point
$P_c$ decreases as the temperature increases. This result is consistent with
experimental observation \cite{Liao}. However, for the experimental
temperature $t=175$ nK (about $0.0267 \varepsilon_b$), our theoretical
intersection point is smaller than the experimentally estimated value. This
discrepancy originates merely in defining the edge of the cloud in the
experiment where the phase boundaries are no longer sharp lines \cite{note}.
We also find that increasing temperature results in shrinking the
fully-paired phase and populating the vacuum. The existence of these quantum
phases at low temperatures is a manifestation of different TLL phases, i.e.,
a two-component TLL of FLLO-like states and a TLL of hard-core bosons and a
TLL of unpaired fermions \cite{Zhao}, where the low-lying excitations are
close to the Fermi points.

\begin{figure}[t]
{{\includegraphics [width=1.0\linewidth]{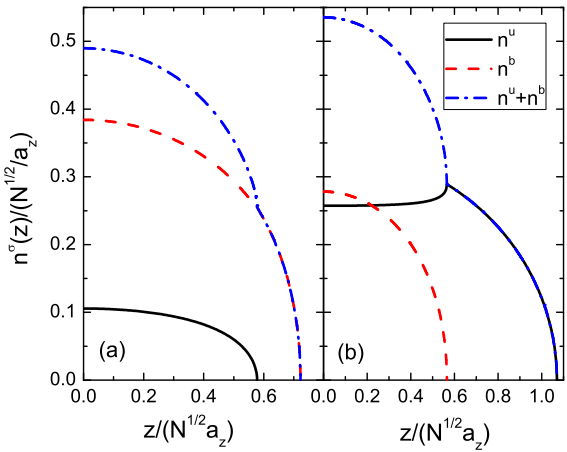}}}
\caption{The zero-temperature density distribution profiles
for trapped fermions with ${Na_{1D}^{2}}/{a_{z}^{2}}=0.2$ for (a) $P=10\%$ 
and (b) $P=50\%$. Here we $n^{u}=n_{\uparrow }-n_{\downarrow
}$ and $n^{b}=n_{\downarrow }$.}
\label{fig:density}
\end{figure}

The FFLO-like state behaves like a mixture of composite fermions with mass $%
2m$ and excess fermions with mass $m$ \cite{Guan-Ho}. The density of state
changes critically as the driving parameter chemical potential is varied
across the phase boundaries. Fig. \ref{fig:density} shows the density
distributions 
for ${Na^2_{1D}}/{a_z^2}=0.2$ at zero temperature and for two different
values of polarization. For small polarization ($10\%$), an FFLO-like state
coexists in the trap center with a BCS-like state at the edges; while for
large polarization ($50\%$), the FFLO-like state lies in the trap center
with a fully polarized normal state at the edges. We also find the FFLO-like
state occupies the whole trap for the critical polarization ($16\%$), which
is consistent with the experimental observation \cite{Liao}.
We now demonstrate how to map out the zero-temperature
phase diagram and quantum criticality from finite temperature density and
compressibility profiles of the trapped gas.

\begin{figure}[tbp]
\includegraphics[width=1.1\linewidth]{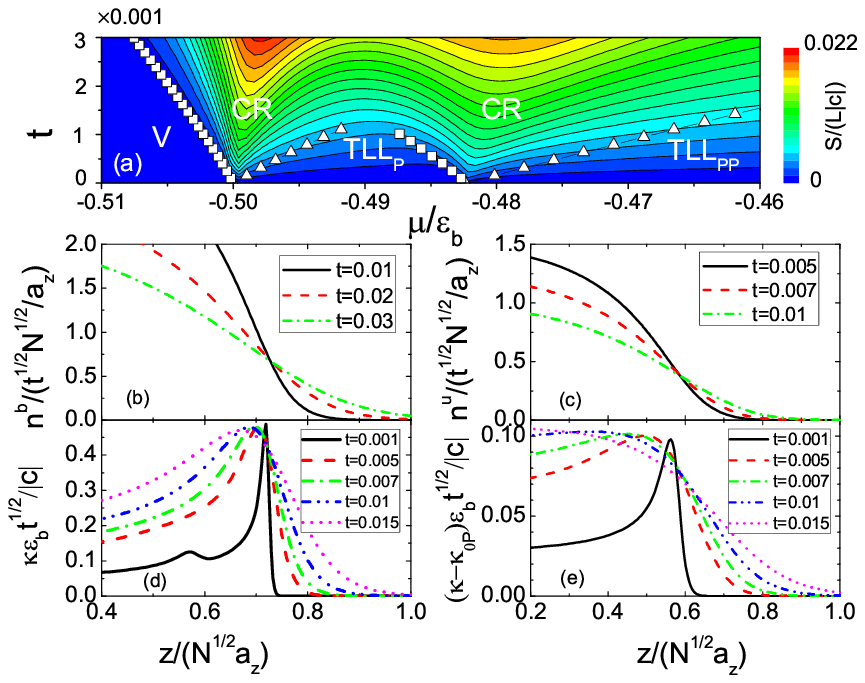}
\caption{Quantum criticality for low polarization. (a)
contour plot of entropy in the $t-\mu$ plane where $t=T/\varepsilon_b$. 
Diamonds and triangles indicate crossover temperatures
separating the quantum critical regimes from vacuum, single component $TLL_P$
of paired fermions and two-component $TLL_{PP}$ of FLLO-like states. The
density profiles demonstrate how to locate the critical points (b) $\mu_{c2}$ and (c) $\mu_{c4}$ 
and thus map out their phase boundaries. The corresponding 
compressibility curves (d) and (e)  intersect at the critical points after a
subtraction of the background compressibility. 
}
\label{fig:QC-1}
\end{figure}

\begin{figure}[tbp]
\includegraphics[width=1.1\linewidth]{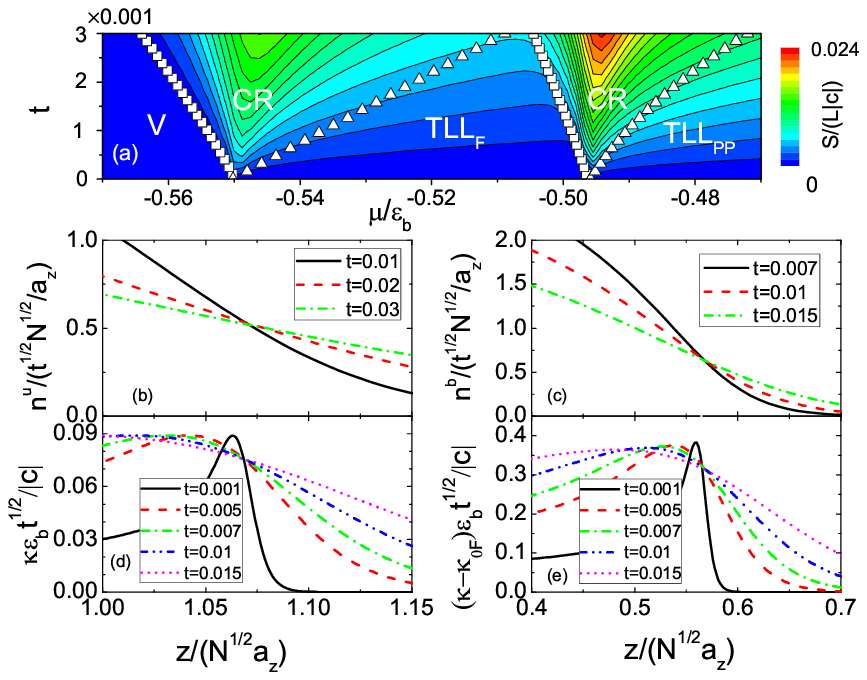}
\caption{Quantum criticality for high polarization. (a)
contour plot of entropy in the $t-\mu$ plane where $t=T/\varepsilon_b$. 
Diamonds and triangles indicate crossover temperatures
separating the quantum critical regimes from vacuum, single component $TLL_F$
of unpairs and two-component $TLL_{PP}$ of FLLO-like states. The density 
profiles demonstrate how to locate the critical points (b) $\mu_{c1}$ and (c) $\mu_{c3}$  
and thus map out their phase boundaries. The corresponding compressibility
curves (d) and (e)  intersect at the critical points after a subtraction of the 
background compressibility. 
}
\label{fig:QC-2}
\end{figure}

\noindent
\textit{Quantum criticality in a harmonic trap.-}
The system exhibits universal scaling behaviour in the vicinity of the critical points 
as a result of the many-body effects. 
The thermodynamic functions of the homogeneous gas
can be cast into a universal scaling form \cite{QC-book,Zhou-Ho,Mueller2}. 
E.g., the density and compressibility scale as 
\begin{eqnarray}
n(\mu, T) &=& n_{0}+ T^{\frac{d}{z}+1-\frac{1}{\nu z}} 
\mathcal{G}\left(\frac{\mu-\mu_{c}}{T^{\frac{1}{\nu z}}}\right), \\
\kappa(\mu, T) &=& \kappa_0+ T^{\frac{d}{z}+1 -\frac{2}{\nu z}}
\mathcal{F}\left(\frac{\mu-\mu_{c}}{T^{ \frac{1}{\nu z}}}\right). 
\end{eqnarray}
Here the dimensionality $d=1$, dynamical critical exponent $z=2$ and correlation length 
exponent $\nu =1/2$ for the 1D strongly attractive Fermi gas \cite{Guan-Ho}. 
The universal scaling functions $\mathcal{G}(x)$ and $\mathcal{F}(x)$  
describe quantum and thermal fluctuations  at  quantum criticality  after 
subtraction of the background density $n_0$ and compressibility $\kappa_0$.

For small polarization $P<P_c$, the chemical potential passes the
lower critical point $\mu_{c2}=-\frac{1}{2}$ from the vacuum into the fully
paired phase then passes the upper critical point 
$\mu_{c4}\approx -\frac{h}{2}+\frac{4}{3\pi}(1-h)^{\frac{3}{2}}+\frac{3}{2\pi^2}(1-h)^2$ 
from the fully
paired phase into the FFLO-like phase. At finite temperatures, contour plots
of the entropy clearly indicate universal critical behavior near the
critical points, see Fig.~\ref{fig:QC-1}(a). The typical $V$-shape crossover
temperatures separate the quantum critical regimes of the hard-core bosonic 
$TLL_P$ phase and the two-component FLLO-like $TLL_{PP}$ phase near the
critical points $\mu_{c2}$ and $\mu_{c4}$, respectively. 
Except for the left-most line separating the vacuum from the
critical regime, all of the lines can be determined by the breakdown of the
TLL, i.e., when the entropy curves deviate from linear $T$-dependent
relations for fixed values of $\mu$ and $h$. These boundaries indicate a
crossover from linear dispersion into nonrelativistic dispersion rather than
phase transitions at finite $T$. 

For large polarization $P>P_c$,
the chemical potential passes the lower critical point $\mu_{c1}=-{h}/{2}$
from the vacuum into the fully unpaired phase then passes the upper critical
point 
$\mu_{c3}\approx-\frac{1}{2}\left(1-\frac{2}{3\pi}(h-1)^{\frac{3}{2}}-\frac{2}{3\pi^2}(h-1)^2\right)$ 
from the fully unpaired phase into the
FFLO-like phase, see Fig. \ref{fig:QC-2}(a). Here the typical $V$-shape
crossover temperatures indicate quantum criticality of the largely polarized
Fermi gas near the two critical points $\mu_{c1}$ and $\mu_{c3}$,
respectively.

The zero temperature phase boundary separating the vacuum from the fully
paired phases can be mapped out from the universal scaling behaviour of the
pair density curves for different temperatures, which intersect at the
critical point $\mu_{c2}$, see Fig. \ref{fig:QC-1}(b). Similarly the phase
boundary separating the fully paired phase from the FFLO-like phase can be
mapped out from the density profiles of unpaired fermions in the trapped gas
at different temperatures, i.e., the unpaired density curves for different
temperatures intersect at the critical point $\mu_{c4}$, 
see Fig. \ref{fig:QC-1}(c). Here we have chosen the densities without background near the
critical points for practical purposes. We see that the density curves for
temperatures below $0.02\varepsilon_b$ well intersect at the critical point.
Accordingly, the compressibilities at low temperatures also well intersect 
at the critical points, see Fig. \ref{fig:QC-1}(d,e).  Here the background compressibility near 
$\mu_{c4}$ is $\kappa_{0P}=-\frac{2|c|}{\varepsilon_b
\protect\sqrt{\pi t}}f_{-\frac{1}{2}}^b\left( 1+\frac{3\sqrt{
t}}{2\protect\sqrt{\pi}}f_{\frac{1}{2}}^b\right)$. 
The results demonstrate that
universal scaling behaviour persists in the trapped cloud of 
atoms \cite{footnote}. As the temperature decreases, we see that the compressibility
evolves a round peak sitting in the phase of the higher density of state. It 
diverges at zero temperature.

For the high polarization case, the density profiles of unpaired and paired
atoms can be used to map out the phase boundaries $\mu_{c1}$ ($V\to F$ ) and
$\mu_{c3}$ ($F\to PP$), respectively, see Fig. \ref{fig:QC-2}(b,c).
Moreover, we show that the corresponding compressibility curves of the trapped gas at low
temperatures intersect at the critical points $\mu_{c1}$ and $\mu_{c3}$, 
 see Fig. \ref{fig:QC-1}(d,e). Near $\mu_{c3}$ the background compressibility is 
 $\kappa_{0F}=-\frac{|c|}{2\sqrt{2} \varepsilon_b \sqrt{\pi t}}f_{-\frac{1}{2}}^u$.
In a similar fashion,
the universal scaling for other thermodynamic quantities such as the specific
heat and magnetization are testable through the quantum criticality of the
trapped gas.
This provides a reliable way to determine quantum phase diagrams and test
universal scaling theory and TLL physics in 1D quantum gases of cold atoms.

\begin{acknowledgments}
This work is supported by the Australian Research Council and by NSF of
China and 973-projects of MOST. The authors thank T.-L. Ho and E. J. Mueller for
stimulating discussions.
\end{acknowledgments}

$\ddagger$ E-mail address: xwe105@physics.anu.edu.au

\end{document}